\documentclass[reprint]{revtex4}
\usepackage{graphicx}
\usepackage{color}
\usepackage{amsmath}
\usepackage{multirow}
\usepackage{amssymb}
\usepackage{amsbsy}
\usepackage{hyperref}
\usepackage{subfigure}
\usepackage{latexsym}
\usepackage[toc,page]{appendix}
\usepackage{setspace}
\usepackage{mathrsfs}
\usepackage{lipsum}
\usepackage{bm}

\def\be{\begin{equation}}
\def\besub{\begin{subequations}}
\def\eesub{\end{subequations}}
\def\ee{\end{equation}}
\def\bea{\begin{eqnarray}}
\def\eea{\end{eqnarray}}

\def\be{\begin{equation}}
\def\ee{\end{equation}}

\def\bwd{\begin{widetext}}
\def\ewd{\end{widetext}}

\def\bp{{\bf p}}

\newcommand{\bsf}[1]{\textsf{\textbf{#1}}}
\DeclareMathOperator{\Tr}{Tr}
\DeclareMathAlphabet{\mathscrbf}{OMS}{mdugm}{b}{n}
\begin{document}
\title{Activating membranes}
\author{Ananyo Maitra$^{1\ast}$, Pragya Srivastava$^{2\ast\dagger}$, Madan Rao$^{2,3}$ and Sriram Ramaswamy$^{4,}$}
\affiliation{Indian Institute of Science, Bangalore 560012, India\\$^2$Raman Research Institute, C.V. Raman Avenue, Bangalore 560 080,
India\\$^3$National Centre for Biological Sciences (TIFR), Bellary Road,
Bangalore 560 065, India\\$^4$TIFR Centre for Interdisciplinary Sciences, Hyderabad 500075, India}


\begin{abstract}
We present a general dynamical theory of a membrane coupled to an actin cortex
containing
polymerizing filaments with active stresses and currents, and demonstrate that
active
membrane dynamics
[Phys. Rev. Lett \textbf{84}, 3494 (2000)] and spontaneous shape oscillations emerge from this description. We
also consider membrane instabilities and patterns induced by the presence of
filaments with polar orientational correlations in the tangent plane of the
membrane. The dynamical features we predict should be seen in a variety of cellular contexts
involving the dynamics of the membrane-cytoskeleton composite and cytoskeletal extracts coupled to synthetic vesicles.
\end{abstract}
\date{\today} 
\maketitle
The plasma membrane of a living cell displays striking dynamical structures
in the form of growing tubules, ruffles, ridges, and spontaneously generated
waves \cite{mogliner_review,Taiwan, Sheetz}. The generality of these
observations prompts us to search for a minimal physical description,
independent of system-specific detail, unlike \cite{Kruse,Gholami}.
We show that the essential mechanism lies in the interaction of the membrane
with the cytoskeleton, driven by molecular motors and ATP, which can be viewed
as a fluid containing orientable, self-driven filaments \cite{review}. 
Our work is significantly different from a recent paper on membrane waves
driven by actin and myosin \cite{gov}, as we explain in the paper. Our
predictions are general, and testable in extracts and artificial settings as
well.
                                                                    
Our main results: (i) Active membrane dynamics \cite{old,RamTonPro,Pumplong}
emerges naturally from a complete hydrodynamic theory of a membrane forced by a
fluid containing orientable motile elements carrying active stresses. The
membrane acquires a tension from the intrinsic stresses on the filaments and a
sustained normal velocity from their nonequilibrium directed motion. (ii) Within
a mode-truncated description, we find an instability to a spontaneously
oscillating state. (iii) Including an in-plane polar orientation field in the
membrane gives rise to height bands just past the onset of spontaneous
alignment, travelling instabilities deep in the ordered phase, with growth rate
$\sim q_x^{1/2}q_y${, where $x$ is the direction of mean ordering of the
filaments,} for small in-plane wavevector $(q_x,q_y)$, and possibly tubules or
ridges in a regime where the polarization focuses onto points or lines.

\begin{figure}
   \begin{center}
    \subfigure{\includegraphics[height=2.5in]{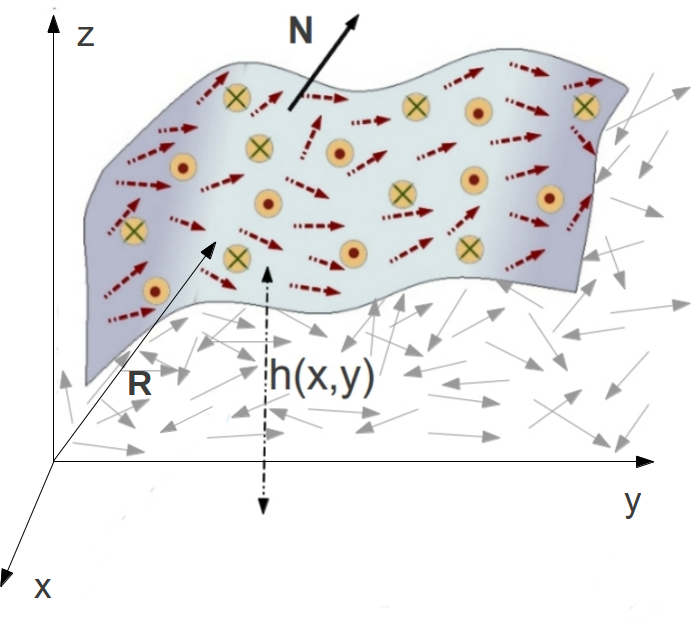}}
     \end{center}
  \caption{Schematic diagram of a membrane in an active fluid. The membrane is depicted in Monge gauge: ${\bf R}=(x, y, h(x, y, t))$ where $h(x, y, t)$ is the height above points $(x, y)$ on a reference plane. The `signed' species $\psi$ is represented by circles with dots (crosses) denoting parallel (antiparallel) alignment with respect to the outward membrane normal ${\bf N}$.
Dashed arrows on the membrane are the in-plane polar filaments, whose concentration is $c$, and continuous arrows denote the bulk active orientable fluid.}
  \label{fig1}
  \end{figure}

We now construct the dynamics of a fluid membrane coupled to a bulk solvent 
\cite{Sumithra, Niladri}
containing active orientable particles \cite{conc} described by a vector order
parameter ${\bf P}({\bf r})$ and a nematic order parameter $\bsf{Q}({\bf r})$,
as functions of 3-dimensional position ${\bf r}$. The membrane
conformation ${\bf R}(\vec{u})$ is parametrized by $\vec{u} = (u_1,u_2)${, where $\vec{u}$ is a two dimensional position vector labelling points in the membrane}. { The local membrane velocity is denoted by ${\bf V}_m(\vec{u},t)$. If we impose that all surface points labelled by $\vec{u}$ retain their coordinates (i.e. we use a convected coordinate system \cite{diffgeom, Hu}), ${\bf V}_m\equiv \partial_t{\bf R}$. We will work with this choice in the present paper. In \cite{supp} we present the general equations for our model.} We
denote by $\psi$ the ``signed'' concentration of a species living on or closely
associated to the membrane (Fig. \ref{fig1}). That is, each particle of this
species has a
vectorial orientation, whose axis is taken to lie along the membrane normal, and
is counted as $+$ ($-$) for parallel (antiparallel) alignment. $\psi$ could
represent \cite{old} actin polymerization nucleators, asymmetric membrane
proteins or ion channels, or an internal state coupling to local curvature
\cite{Hsuan-Yi}. We denote by $c$ the concentration of polar filaments
restricted to the immediate vicinity of the membrane. In the cellular context, this
represents tangential actin, whose presence has been persuasively argued for in
recent studies on membrane composition and trafficking \cite{Mayor}.

In the absence of flow and activity, the system relaxes to equilibrium governed by a free-energy functional $F[{\bf P}, \bsf{Q}, {\bf R}, \psi, c]
=\int d^3r
(f_b+\int_u f_m)$ with contributions $f_b$ from the cytoplasm and $f_m$ at 
the membrane. Here $\int_u...\equiv\int d^2 u g^{1/2}\delta ({\bf r}-{\bf
R})...$, $g$ being the determinant of the membrane metric. $f_b=(a_1/2)
P^2+(a_2/2) \bsf{Q}\colon \bsf{Q}$ controls the relaxation of the order parameter
fields in the bulk. The contributions from the
degrees of freedom associated with the membrane
are contained in $f_m=f_c+f_{R}+f_{R-c}+f_{op}$. Here $f_c$ describes the cost
of concentration fluctuations of { both the in-plane polar filaments and the
signed species}, and $f_R=(\kappa/2)(\mbox{Tr}\,\bsf{K})^2$
penalizes deformations of the membrane,  where $\bsf{K}$ is the curvature
tensor \cite{Cai}. $f_{R-c}=\Upsilon(c)\psi{\Tr \bsf{K}}$
couples the signed density field and the local mean curvature \cite
{RamTonPro}. 
Lastly,
\begin{eqnarray}
f_{op}&=&w\bsf{Q}_u\colon {\bf NN}-\Lambda \psi p_n-w_2 p_n {\Tr
\bsf{K}}+a_c\vec{p_t}^2\nonumber\\&&+\alpha c\mathcal{D}_u\cdot
\vec{p_t}+\kappa_p \psi\bsf{K}\colon ({\bf \mathcal{D}}_u \vec{p_t})+\kappa_t
\psi \vec{p_t}\vec{p_t}\colon\bsf{K}
\label{free}
\end{eqnarray}
is the free-energy associated with orientational order at the
membrane: ${\bf P}_u(\vec{u})={\bf P}({\bf r}={\bf R}(\vec{u}))$ and
$\bsf{Q}_u(\vec{u})=\bsf{Q}({\bf r}={\bf R}(\vec{u}))$, parametrised by
$\vec{u}$. $\mathcal{D}_u$ is the covariant
derivative on the membrane \cite{diffgeom} and ${\bf
N}(\vec{u})$ is the membrane normal. We decompose ${\bf P}_u$ into its normal component $p_n={\bf P}_u\cdot{\bf N}$
and the tangent plane vector
$\vec{p_t}={\bf e}^u\cdot{\bf P}_u$, where the projector ${\bf e}^u\equiv({\bf
e}^1(\vec{u}), {\bf e}^2(\vec{u}))=(\partial_{u_1}{\bf R}, \partial_{u_2}{\bf
R})$. The local polarity $\psi$ of the membrane favours one direction of ${\bf
P}$ through $\Lambda$, while $w$, depending on its sign, softly anchors the
filaments parallel or perpendicular to the membrane \cite{Norio}.
$w_2$, $\kappa_p$ and $\kappa_t$ couple orientation to curvature
\cite{Nelson},
and $a_c$ controls the
orientational free energy of membrane-associated tangential polar
filaments (hereafter, ``horizontal filaments").
The coefficient $\alpha$ governs local spontaneous splay in response to
polar filament concentration\cite{Kung, kripa}. 

In the presence of active processes, the membrane, treated as a permeable fluid
film \cite{Cai}, has a local velocity
\begin{equation}
{ {\bf V}_m}= [{\bf V}+v_0 {\bf P}+\zeta \nabla\cdot({\bf PP})]|_{{\bf r}={\bf R}}-\mu_p g^{-1/2} \frac{\delta F}{\delta {\bf R}}
\label{conformation}
\end{equation}
where ${\bf V}$ is the three dimensional hydrodynamic velocity. The second
and third terms in the square bracket \eqref{conformation}, forbidden in a
passive system, arise as follows \cite{review}: free
energy is dissipated at a rate $\mathscr{R}\Delta\mu$,
where $\mathscr{R}$ is the reaction rate and $\Delta \mu$ the chemical potential
difference between the fuel (e.g., ATP) and its reaction products. Let us treat
${ {\bf V}_m}$ and $\mathscr{R}$ as
fluxes \cite{deGroot, Curie}, with corresponding forces $\delta F / \delta {\bf R}$ and $\Delta \mu$.
To first order in gradients, ${\bf P}$ and $\nabla\cdot ({\bf PP})$,
measuring
local polarity, contribute terms of the form $\zeta_1 {\bf P} \cdot
\delta F / \delta {\bf R}$ and $\zeta_2 \nabla\cdot ({\bf PP})\cdot \delta F /
\delta {\bf R}$ to $\mathscr{R}$, where the independent kinetic coefficients
$\zeta_1$ and $\zeta_2$ vanish for an impermeable membrane, as does $\mu_p$. The
symmetry of dissipative Onsager coefficients then implies terms
$\zeta_1\Delta\mu{\bf P}\equiv v_0{\bf P}$ and $\zeta_2\Delta\mu \nabla\cdot
({\bf PP})\equiv \zeta\nabla\cdot ({\bf PP})$ in the ${ {\bf V}_m}$
equation. In the cellular context $v_0$ is the scale of the drift speed of the
membrane arising from filament polymerization \cite{Carlier}. 

The signed density field $\psi$ has a dynamics given by 
\begin{equation}
\mathcal{D}_t \psi=-\mathcal{D}_u\cdot\vec{J}-k_2 \psi+k_1
\label{psidyn}
\end{equation}
where $\mathcal{D}_t$ is the covariant time derivative, and $k_1$, $k_2$ are
rates of association and dissociation to the membrane. The
current 
\begin{equation}
\label{Jdef}
\vec{J}=\vec{J}_0\equiv-\psi{\bf e}^u\cdot\partial_t {\bf R}-D\left[
\psi \mathcal{D}_u\left(g^{-1/2}\frac{\delta F}{\delta\psi}\right)\right],
\end{equation}
contains drift and diffusion; the diffusivity $D$ can include active
contributions.  

The Stokesian hydrodynamic velocity field ${\bf V}({\bf
r})=\int_{\bf r'} \bsf{H}({\bf r}-{\bf r'}) \cdot \mathcal{F}({\bf r'})$,
where $\bsf{H}({\bf r'})$ is the Oseen tensor \cite{oseen}. The force
density $\mathcal{F}({\bf r},t)= \nabla \cdot \bm{\sigma}^\bsf{Q}+\int_u \delta
F/\delta {\bf R}$, where the order parameter stress
\cite{degen} $\bm{\sigma}^\bsf{Q}$ has an active contribution of the form
$\zeta_Q \bsf{Q}$ \cite{PP,Nir}.

The order parameters have standard equations of motion \cite{review} $\dot{\bf
P}=-\Gamma_p\delta F/\delta {\bf P}$ and $\dot{\bsf{Q}}=\lambda_0
\bsf{S}-\Gamma_Q\delta F/\delta \bsf{Q}$, where $\bsf{S}$ is the
symmetrised velocity gradient tensor.

The local dynamics of the membrane on scales small compared to the whole cell
can be understood in the Monge gauge (Fig. \ref{fig1}), ${\bf R}=({\bf x}, h({\bf x}, t))$,
$g=1+(\nabla_{\perp}h)^2$ and ${\bf N}=g^{-1/2}(-\nabla_{\perp}h, 1)$, where $h({\bf x})$
is the height of the membrane above a point ${\bf x}$ on the reference plane. We first concentrate on the coupled dynamics of $h$ and
$\psi$ in an isotropic bulk phase with
negligible $c$. On long time-scales the order parameters relax to values
governed by their coupling to the
membrane which, to lowest order in gradients, are
\begin{subequations}
\begin{equation}
{\bf P}_u=\psi\frac{\Lambda}{a_1}{\bf N},
\label{Prel}
\end{equation}
\begin{equation}
\bsf{Q}_u=\frac{\Gamma_Q\lambda_0}{a_2} \bsf{S}|_{{\bf r}={\bf R}}-\frac{w}{a_2}({\bf NN}-{\bf I}/3). 
\label{Qrel}
\end{equation}
\end{subequations}
{ Using these expressions for ${\bf P}$ and $\bsf{Q}$ in \eqref{conformation}, \eqref{psidyn} and \eqref{Jdef} }leads to the coupled equations
\begin{widetext}
\begin{subequations}
\begin{equation}
\partial_t h=g^{1/2}[\tilde{v}\psi-\mu_p\frac{\delta F}{\delta h}]-\int_{\bf
q}e^{i{\bf q}_{\perp}\cdot{\bf
x}}\frac{1}{4\tilde{\eta}q_{\perp}}(\gamma_{act}q_{\perp}^2h_{\bf q}+\frac{\delta
F}{\delta h})
\label{actpump1}
\end{equation}
\begin{equation}
\partial_t \psi= \tilde{v} \nabla_{\perp}\cdot(g^{-1/2}\psi^2\nabla_{\perp}
h)+D\nabla_{\perp}\cdot[\psi\nabla_{\perp}(g^{-1/2}\frac{\delta F}{\delta
\psi})]-k_2\psi+k_1
\label{actpump2}
\end{equation}
\label{actpump}
\end{subequations}
\end{widetext}
in which the active stress contributes a tension $\gamma_{act}=
-\zeta_Qw/a_2$ (see \cite{Norio} for a one-dimensional analogue) and, as in \cite{Yhat}, a modified viscosity
$\tilde{\eta}=\eta+(-\lambda_0a_2+\zeta_Q)\Gamma_Q\lambda_0/a_2$. { We have
not included the effects of the active term with the coefficient $\zeta$
defined in \eqref{conformation}. The lowest order term it contribues to the height
equation is $\nabla_\perp^2 h$, i.e. an active non-hydrodynamic tension. Its more
crucial consequences are examined later on in the paper} Note that the
active modification of { the hydrodynamic} tension is missing in \cite{Pumplong}, where the
force dipoles are taken to be situated at the reference rather than the
actual location of the membrane. Local polarity leads to the propulsion of the
membrane at a rate $\psi\tilde{v}=\psi v_0
\Lambda/a_1$. $F$ in \eqref{actpump1} and \eqref{actpump2} is the original
free-energy functional with ${\bf P}$ and $\bsf{Q}$ eliminated in favour of
$\psi$ and $h$ via \eqref{Prel} and \eqref{Qrel} respectively. The first term
on the right-hand side in \eqref{actpump2} arises kinematically, due to
the change of density of an in-membrane species resulting from a change in the conformation of the fluid film\cite{Cai}.
Setting $k_1$ and $k_2$ to zero to specialise to the case of a { zero mean} conserved
species, eqns. \eqref{actpump1} and \eqref{actpump2} take precisely the form
presented \cite{RamTonPro} on general grounds for an active membrane
\cite{pt1a}. This establishes one of our main results: a membrane in an
active fluid is an active membrane, propelled by polar activity --
polymerization, in the context of actomyosin -- with a tension from
contractility. Equations \eqref{conformation} - \eqref{actpump} with suitable
boundary conditions can be applied to cell membranes or reconstituted systems. 


The terms proportional to $\tilde{v}$ in \eqref{actpump}
constitute an excitatory-inhibitory pair which, we now show, leads to sustained spontaneous
oscillations in a regime of parameter space. We work in one dimension, retaining only the smallest
wavenumber and only one nonlinear term:
$\nabla_{\perp}\cdot(\psi^2\nabla_{\perp} h)$ in \eqref{actpump2}. The resulting
coupled ODEs \cite{supp} upon re-scaling and defining new constants describe a
generalized Van der Pol oscillator with a linear damping and
cubic nonlinearities: 
\begin{equation}
\ddot{\phi} +\phi+s_1\dot{\phi}+s_2 \phi^2\dot{\phi}+s_3 \phi
(\dot{\phi})^2 +s_4 \phi^3=0
\end{equation}
which has been shown \cite{rand} to have a limit cycle if
$\mbox{sgn}(s_1s_2)=-1$. We provide further details regarding the mode truncation and present a representative phase portrait in the supplement \cite{supp}. 

Note that wave-like dispersion relations as in \cite{gov_gop, gov} are distinct
from the experimentally observed membrane waves
\cite{Taiwan,Sheetz}, which are not a response to an external perturbation,
but are self-generated, in a manner consistent with our findings from the
truncated model. Moreover, the wave-speed in our theory is set solely by the
normal drift speed, not by free-energy couplings.

Our analysis of \eqref{actpump}
suggests an explanation for the experimentally observed waves \cite{Taiwan,
Sheetz} on the lamellipodium \cite{Small2002} of a crawling or spreading cell,
whose leading edge should be viewed as an actively moving one-dimensional
membrane. We expect similar waves on the surface of self-propelled drops
\cite{Cates} e.g., in parameter regimes corresponding to the
instability discussed in \cite{RamTonPro}, which arises here if $\tilde{v}
\Upsilon > 0$. In the case we present here and in \cite{supp} actin
polymerization, not contractility, is the proximate cause of the membrane waves
\cite{Sheetz}. Note: even without permeability, a local normal velocity at the
membrane, proportional to $|q_{\perp}|\psi_{\mathbf{q}}$, which can be shown to
arise from an active contractile stress, can generate spontaneous waves with
dispersion $\omega^2\sim q^3$ \cite{Pumplong, supp}.

Now we examine another case of importance to cell biology, in
which a distinct population of filaments, lying in the vicinity
of the membrane and disposed parallel to it, are present at sufficient
concentration for their dynamics to be slow and therefore relevant on the
timescale of interest to this work. { This is motivated by experimental
studies \cite{Mayor} of the nanoclustering of cell-surface molecules, whose
anomalous statistical properties are naturally accounted for as arising from
active transport mediated by a new class of ``horizontal actin filaments''. 
We study the effect of such tangential active orientable filaments on
membrane fluctuations, and make predictions which can be tested in future
experiments. For this, we introduce a separate dynamical equation} for the polar
order parameter ${\bf P}_u$ at the
membrane \cite{Cai}: 
{
\begin{equation}
\mathcal{D}_t {\bf P}_u-({\bf e}^u\cdot\partial_t{\bf R})\cdot\mathcal{D}_u{\bf P}_u+v_p\vec{p_t}\cdot\mathcal{D}_u{\bf
P}_u=-g^{-1/2}\Gamma_p\delta F/\delta {\bf P}_u
\label{Pudyn}
\end{equation}
}
where $v_p$ is the self-propulsion velocity \cite{Propulsion}. In the following
treatment, for simplicity, we replace hydrodynamic damping by local friction
with respect to a fixed background medium. We assume the horizontal filaments
are close to an ordering transition
but take the normal component $p_n$ to relax
rapidly. To the lowest order in gradients, \eqref{Prel} implies $p_n=(\Lambda/a_1) \psi$. The equation for the tangential component is
{
\begin{equation}
\mathcal{D}_t \vec{p_t}-{\bf e}^u\cdot[({\bf e}^u\cdot\partial_t{\bf R})\cdot\mathcal{D}_u{\bf P}_u]+v_p{\bf e}^u\cdot(\vec{p_t}\cdot\mathcal{D}_u{\bf
P}_u)=-\frac{1}{\sqrt{g}}\Gamma_p{\bf e}^u\cdot\frac{\delta
F}{\delta {\bf P}_u}+(\mathcal{D}_t{\bf e}^u)\cdot{\bf P}_u,
\label{transverse}
\end{equation}
}

Conservation of horizontal filaments implies 
\begin{subequations}
\begin{equation}
\mathcal{D}_t c=-\mathcal{D}_u\cdot(c\vec{v}_c)+D_c
\mathcal{D}_u\cdot[c\mathcal{D}_u(g^{-1/2}\frac{\delta F}{\delta c})],
\label{cdyn}
\end{equation}
\begin{equation}
\vec{v}_c=-{\bf e}^u\cdot\partial_t {\bf R}+v_1 \vec{p_t}. 
\label{ccurr}
\end{equation}
\label{concfull}
\end{subequations}
The membrane conformation is given by \eqref{conformation}, while the current of
$\psi$ in \eqref{psidyn} is modified to $\vec{J}=\vec{J}_0+v_{\psi}\vec{p_t}$.
$v_1$ and $v_{\psi}$ are active polar velocity parameters, independent of each other and of $v_p$ in \eqref{Pudyn}.

Eqs. \eqref{conformation}, \eqref{psidyn}, \eqref{transverse} and \eqref{concfull} are a
formally complete description of the dynamics of a membrane endowed
with in-plane polar orientational order and signed species coupled to
active `horizontal' and `vertical' filaments. A complete exploration of the
range of behaviors of
this system requires a numerical study. We limit ourselves here
to a linear stability analysis about the isotropic and in-plane ordered states
in a steadily moving membrane which is flat on average. This is the regime in
which dynamics of $\psi$ is fast and relaxes to a steady state value
$\psi_0=k_1/k_2$, \eqref{psidyn}. As $k_2^{-1}\sim 0.1-1$s
\cite{onoff} our assumption is justified if we are looking at the dynamics on
timescales greater than $1$s. We re-scale our equations so that $\psi_0=1$.

The coupled equations of the height field $h$, the in-plane component of the
polar order parameter ${\bf p}$ and concentration $c$, to leading order in
gradients, are
\begin{widetext}
\begin{subequations}
\begin{equation}
\partial_t h=\tilde{v}(c)
+\frac{\zeta}{a_1}\nabla_{\perp}\cdot[\Lambda(c){\bf p}]+\mu_p[\Sigma\nabla_{\perp}^2h-\kappa\nabla_{\perp}^4h 
+\nabla_{\perp}^2\Upsilon(c)-\kappa_p\nabla_{\perp}^2\nabla_{\perp}\cdot{\bf p}-\kappa_t\nabla_{\perp}\nabla_{\perp}\colon{\bf pp}],
\label{Mongeh}
\end{equation}
\begin{equation}
\partial_t {\bf p}=-v_p\bp\cdot\nabla_{\perp}\bp+\frac{\tilde{v}(c)v_0}{a_1}\nabla_{\perp}\Lambda(c)+\Gamma_p[-\tilde{A}{\bf
p} 
+\kappa_p\nabla_{\perp}\nabla_{\perp}^2
h+\alpha\nabla_{\perp} c-\kappa_t{\bf p}\cdot\nabla_{\perp}\nabla_{\perp}h]+D_p\nabla_{\perp}^2 {\bf p}
\label{planeP}
\end{equation}
\begin{equation}
\partial_t c=\nabla_{\perp}\cdot[c(v_1 {\bf
p}+\tilde{v}(c)\nabla_{\perp}h)]+D_c\nabla_{\perp}^2 c, 
\label{concmonge}
\end{equation}
\label{coupledpch}
\end{subequations}
\end{widetext}
where $\nabla_{\perp}^2 \Upsilon(c)$ arises from the free energy contribution
$f_{R-c}$. $\Sigma$ in \eqref{Mongeh} is an active tension \cite{RamTonPro},
arising, for example, via an interplay of the active polymerization and the
polar anchoring modelled by the free-energy cost $-\int_u w_2 p_n{\Tr \bsf{K}}$
in \eqref{free}.
This coupling generates a term of the form $\nabla_{\perp}^2h$ in the $p_n$
equation and, therefore, because of propulsion, an effective tension in the $h$ equation.
\eqref{planeP} was obtained by projecting \eqref{transverse} onto the reference
horizontal, with $\tilde{A}=(a_c+a_1)$. The term with coefficient $D_p$ arises
from the Frank elasticity of the polar filaments. Note the absence of a term
proportional to $\nabla_{\perp}h$ in \eqref{planeP}, a consequence of
three-dimensional rotation invariance \cite{Sumithra}. The propulsive velocity
$\tilde{v}$ in \eqref{coupledpch} is taken to depend only on $c$ as $\psi$ has
been set to a constant value. We turn next to some original instability
mechanisms emerging from \eqref{coupledpch}. 


First
consider the case of large positive $\tilde{A}$. ${\bf p}$ is then deep in
the isotropic phase, and can thus be eliminated in favour of $h$ and $c$ on
timescales long compared to its finite relaxation time. The resulting equations
are then those of \cite{RamTonPro}, with a modified diffusivity $D_c \to
D_c+\alpha v_1/\tilde{A}$. The complete problem, including the dynamics of
${\bf p}$, is characterized by two eigenmodes with relaxation rates $\sim q^0$
and two of order $\sim q^2$, unaffected to leading order in $q$ by the coupling
$\nabla_{\perp}\nabla_{\perp}^2 h$ in the ${\bf p}$ equation. A large enough
negative $\alpha$ leads to an instability with aggregation of $c$ and modulation
of $h$. This picture is borne out by a linear stability analysis in which the
dynamics of ${\bf p}$ is retained \cite{supp}, revealing an eigenvalue of order
$q^2$ that changes sign for sufficiently large negative $\alpha v_1$. The
projection of the corresponding eigenvector onto $h$ grows with increasing
$\tilde{v}$. The underlying process involves the focusing of $\mathbf{p}$ and
hence the concentration, leading, through the $\tilde{v}(c)$ term in
\eqref{Mongeh}, to growth of the height field. Whether the focusing of ${\bf p}$
takes the form of asters or walls, leading respectively to height modulations in
the form of tubules or ridges, requires a numerical calculation. If both
$\tilde{A}$ and $\alpha$ are
negative, an extrapolation of the results of \cite{kripa} would suggest the
formation of ordered modulations of $h$. 

In flocking models \cite{Toner}, just past the onset of the ordered phase of
$\bp$, the coupled dynamics of $c$ and ${\bf p}$ gives rise, through the $c$
dependence of $\tilde{A}$, to a state with travelling bands of concentration
\cite{band}. In the present context where the dynamics takes place on a membrane
this should be accompanied, through $\tilde{v}(c)$ in \eqref{Mongeh}, by a
one-dimensional fore-aft asymmetric modulation of the membrane height.

The coupled dynamics of $\bp$ and $h$ with $c$ fixed shows a distinct class of
modes and instabilities deep in the regime where ${\bf p}$ is ordered. For
$v_p=0$ there is a travelling instability with relaxation rate of transverse
fluctuations $\sim q_yq_x^{1/2}$, if the ordering direction is taken along $x$.
As $v_p$ is increased this crosses over to $\sim q_y^2$ \cite{supp}. Note,
despite the similarity of form with the mode structure of \cite{Sumithra2}, that
the detailed mechanisms are different.

To summarize: we have shown that the equations of motion for an active membrane
\cite{RamTonPro} emerge from the dynamics of an ordinary fluid membrane coupled
to a medium with active, motile filaments. We find that the resulting equations
display spontaneous sustained oscillations driven by the active motion of the
membrane normal to itself which are the natural explanation of membrane waves
\cite{Taiwan, Sheetz}. In addition, when polar ``horizontal filaments"
\cite{Mayor} are included, the coupled dynamics of their concentration and
orientation and the membrane height leads to instabilities towards one- or
two-dimensional modulations, as well as travelling undulations. Deep in the
orientationally ordered
phase of the filaments we find propagating instabilities with singularly
anisotropic dependence on wavevector. Ongoing numerical studies of the long-time
dynamics emerging from these instabilities find a varied range of behaviors
including stable tubules and spatiotemporal chaos \cite{ananyounpub}. Meanwhile,
we look forward to tests of our predictions in actomyosin extracts with ATP and
actin nucleators in contact with model lipid membranes.

\begin{acknowledgments}
AM and SR thank Jean-Fran\c{c}ois Joanny for useful discussions in the early stages of the work. AM, PS and MR thank TCIS, TIFR Hyderabad for hospitality, and SR acknowledges a J.C. Bose fellowship. MR acknowledges a grant from Simons Foundation.
\end{acknowledgments}
$\ast$ Equal contributors

$\dagger$ Present address: Physics department, Syracuse University, Syracuse, NY-13244, USA

\newpage
\section*{Activating Membranes; A. Maitra et al. : Supplementary Material}
In this supplementary, we provide (i) intermediate steps for derivation of the
Eqn. (7) in the main text, starting from the equations describing active membrane
(Eqn.(6) in the main text) along with the phase portrait showing the limit cycle, (ii) explanation of the oscillatory state in an impermeable membrane
and (iii) expressions and plots of dispersion relations for a membrane with an 
in-plane polar order parameter.
{ 
\subsection{Derivation of the membrane equations}

In this section we present the full time-dependent reparametrisation invariant equations for our model. Let $\rho$ be the mass density of the membrane. It follows an equation of motion
\begin{equation}
\mathcal{D}_t\rho+\mathcal{D}_u\cdot(\rho\vec{v})=0
\label{dens}
\end{equation}
where $\vec{v}$ is the membrane velocity field. 
\begin{equation}
{\bf V}_m=\vec{v}\cdot{\bf e}^u+\partial_t{\bf R}
\end{equation}
where contraction on the two-dimensional index of ${\bf e}^u$ is used here
to ``lift'' $\vec{v}$ to three dimensions. In general, the mobility
relating of the membrane with respect to the fluid is anisotropic with the
friction being different normal and tangential to the membrane. In $(2)$ of the
main paper we looked at the special case of an isotropic mobility. The general
expression corresponding to $(2)$ of the main paper is
\begin{equation}
{\bf V}_m={\bf V}|_{{\bf r}={\bf R}}+\bsf{M}\cdot[\tilde{v}_0{\bf P}|_{{\bf r}={\bf R}}+\tilde{\zeta}\nabla\cdot({\bf PP})|_{{\bf r}={\bf R}}-\tilde{\mu_p}g^{-1/2}\frac{\delta F}{\delta {\bf R}}]
\label{memvel}
\end{equation}
where $\bsf{M}$ is an anisotropic mobility. With this formulation one can consider various cases for the membrane-bulk fluid interaction like no-slip in the tangential direction or no-penetration in the normal one. For instance, no-slip condition can be imposed as ${\bf e}^u\cdot{\bf V}_m={\bf e}^u\cdot{\bf V}|_{{\bf r}={\bf R}}$, i.e. the tangential friction is infinite. In this case, $\vec{v}={\bf e}^u\cdot{\bf V}|_{{\bf r}={\bf R}}-{\bf e}^u\cdot\partial_t{\bf R}$. As in \cite{cai}, \eqref{memvel} can be transformed into an equation for normal fluctuations of the membrane $\partial_t{\bf R}\cdot{\bf N}$ and conservation law for density using \eqref{dens} to eliminate $\vec{v}$. 

The dynamical equation for $\psi$ is
\begin{equation}
\mathcal{D}_t\psi=\mathcal{D}_u\cdot(\psi{\bf e}^u\cdot\partial_t{\bf R})-\mathcal{D}_u\cdot(\psi{\bf e}^u\cdot{\bf V}|_{{\bf r}={\bf R}})+D\mathcal{D}_u\cdot\left[\psi \mathcal{D}_u\left(g^{-1/2}\frac{\delta F}{\delta\psi}\right)\right]
\end{equation}
When the tangential filaments are also included, the equation is modified to 
\begin{equation}
\mathcal{D}_t\psi=v_\psi\mathcal{D}_u\cdot(\psi \vec{p}_t)+\mathcal{D}_u\cdot(\psi{\bf e}^u\cdot\partial_t{\bf R})+D\mathcal{D}_u\cdot\left[\psi \mathcal{D}_u\left(g^{-1/2}\frac{\delta F}{\delta\psi}\right)\right]
\label{psieq}
\end{equation}
When these equations are expressed in Monge gauge, the current of $\psi$ has a term proportional to $\psi\dot{h}\nabla_\perp h$ because of metric fluctuations, which gives rise to the kinematic term in eq. (6b) of the main paper.

The dynamical equation for the concentration of tangential filaments
\begin{equation}
\mathcal{D}_tc=v_1\mathcal{D}_u\cdot(c \vec{p}_t)+\mathcal{D}_u\cdot(c{\bf e}^u\cdot\partial_t{\bf R})+D_c\mathcal{D}_u\cdot\left[c \mathcal{D}_u\left(g^{-1/2}\frac{\delta F}{\delta c}\right)\right].
\end{equation} 
The polarisation vector follows the equation
\begin{equation}
\mathcal{D}_t {\bf P}_u-({\bf e}^u\cdot\partial_t{\bf R})\cdot\mathcal{D}_u{\bf P}_u+v_p\vec{p_t}\cdot\mathcal{D}_u{\bf
P}_u=-g^{-1/2}\Gamma_p\delta F/\delta {\bf P}_u
\end{equation}

This constitutes a complete set of equations for our model. 

\subsection{Full expressions for the order parameter equations and the hydrodynamic stress stensor}
{ Consider a suspension of active particles in a fluid. The particle phase velocity ${\bf v}^p$ in such systems is given by
\begin{equation}
{\bf v}^p=\zeta_1{\bf P}+\zeta_2\nabla\cdot{\bsf Q}+{\bf v}+...
\end{equation}
where ${\bf v}$ is the centre of mass velocity, and $\zeta_1$ and $\zeta_2$ are arbitrary activity parameters. The ellipsis denotes the higher order contributions to ${\bf v}^p$ arising from the particle phase stress.}
The three-dimensional polar order parameter has the following equation of motion \cite{Kung2}
\begin{equation}
\partial_t P_i=-v^p_j\partial_jP_i-\lambda^p_{ijk}\partial_kv^p_j+\tilde{\lambda^p}\partial_j\partial_jv^p_i-\Gamma^P_{ij}\frac{\delta F}{\delta P_j}
\end{equation}
$\lambda^p_{ijk}$ describes alignment of the polarisation vector with gradients in velocity. A polarisation vector can also reorient to point along the gradient of strain-rate through the term with coeffcient $\tilde{\lambda^p}$.
\begin{equation}
\lambda^p_{ijk}=\frac{1-\lambda}{2}(\delta_{ij}^TP_k+\delta_{ik}^TP_j+\delta_{jk}^TP_i)+\lambda\frac{P_iP_jP_k}{P^2}
\end{equation}
with 
\begin{equation}
\delta_{ij}^T=\delta_{ij}-\frac{P_iP_j}{P^2}
\end{equation}
and
\begin{equation}
\Gamma^P_{ij}=\Gamma_p\delta_{ij}+\Gamma_1P_iP_j+\Gamma_2(\partial_i P_j+\partial_jP_i)
\end{equation}
to second order in gradients and fields. 

The apolar order parameter is governed by the following dynamical equation \cite{Stark2}:
\begin{equation}
\partial_t Q_{ij}=-v^p_k\partial_k Q_{ij}+\lambda_{ijkl}\partial_lv^p_k
-\Gamma^Q_{ijkl}\left[\frac{\delta F}{\delta Q_{kl}}\right]^{ST}
\end{equation}
We define ${\bsf A}^{ST}$ to be the symmetrised traceless part of any tensor ${\bsf A}$ .  
\begin{multline}
\lambda_{ijkl}=
\frac{\lambda_0}{2}(\delta_{ik}\delta_{jl}+\delta_{jk}\delta_{il}-\frac{2}{3}\delta_{ij}
\delta_{kl})+\frac{1}{2}(\delta_{ik}Q_{jl}-\delta_{il}Q_{jk}+\delta_{jk}Q_{il}
-\delta_{jl}Q_{ik})\\+\frac{\lambda_1}{2}(\delta_{ik}Q_{jl}+\delta_{jk}Q_{il}
+\delta_{jl}Q_{ik}+\delta_{il}Q_{jk}-\frac{4}{3}\delta_{ij}Q_{kl})+\lambda_2Q_{ij}Q_{kl}
\end{multline}
and
\begin{equation}
\Gamma^Q_{ijkl}=\frac{\Gamma_Q}{4}(\delta_{ik}\delta_{jl}+\delta_{il}\delta_{jk} + \delta_{jl}\delta_{ik} +\delta_{jk}\delta_{il}).
\end{equation}

We define the strain rate tensor as $S_{ij}=1/2(\partial_iv_j+\partial_j v_i)-1/3\nabla\cdot{\bf v}\delta_{ij}$. The order parameter equations presented in the main paper are lowest order in gradients and fields versions of these when there is no global orientational order. 

{ The description retaining both P and Q merits some comment. In the phase in which P globally orders along a certain direction, obviously ${\bsf Q}$ has a steady state value $Q_{ij}\propto P_iP_j-(1/3)\delta_{ij}P^2$. The fluctuation equation that one writes down for ${\bsf Q}$ is for deviations away from this state. It is not completely determined by the fluctuations of the ${\bf P}$. Moreover, the polar and apolar order may develop independently, e.g., in the sequence isotropic $\to$ apolar $\to$ polar, in a system in which the individual orientable particles were polar. Also, the presence of a signed species in the membrane can promote local polar ordering even in the parameter regime in which bulk apolar ordering is preferred. Lastly, we neglect terms of the form $P_iP_j$ in the $Q_{ij}$ equation of motion, and $Q_{ij}P_j$ in the $P_i$ equation, which do not affect our conclusions in any essential way as we are expanding about a bulk isotropic fluid. }

We will now explicitly write down the full order-parameter stress tensor $\boldsymbol{\sigma}^{\bsf Q}$ for our model. The active contribution is
\begin{equation}
\boldsymbol{\sigma}^{\bsf Q}_a=\zeta_Q{\bsf Q}+\zeta_P{\bf PP}+\tilde{\zeta_P}[(\nabla {\bf P}+(\nabla {\bf P})^T]
\end{equation}
{The activity  parameters $\zeta_Q$, $\zeta_P$ and $\tilde{\zeta_P}$ are functions of local concentration of the filaments.}
There can also be active contributions to the pressure. However, such terms do not play any role in our model due to incompressibility. The passive contributions to the order-parametr stress to polar and apolar order parameter are, respectively
\begin{equation}
\boldsymbol{\sigma}^{\bsf Q}_{polar}=\partial_i P_k \frac{\partial f}{\partial (\partial_j P_k)}-\lambda^p_{ijk}\frac{\delta F}{\delta P_k}+\tilde{\lambda^p}\left[\partial_i\frac{\delta F}{\delta P_j}+\partial_j\frac{\delta F}{\delta P_i}\right]
\end{equation}
\begin{equation}
\boldsymbol{\sigma}^{\bsf Q}_{apolar}=-(\partial_i Q_{kl}) \frac{\partial f}{\partial
\left(\partial_j Q_{kl}\right)} -\lambda_{klij}\left[\frac{\delta F}{\delta Q_{kl}}\right]^{ST}
\end{equation}
From the above expression of the order parameter stress we see that in the disordered phase the only relevant active contribution is through $\zeta_Q$. 
}

\subsection{Derivation of equation (7) from equation (6)}
Eqn. (6) of the main text in one dimension, if one ignores all non-linearities except $\nabla_{\perp}\cdot(\psi^2\nabla_{\perp}h)$ in (6b), is
\begin{subequations}
\begin{equation}
\partial_t h=\tilde{v}\psi+\Sigma\partial_x^2 h-\mu_p\kappa\partial_x^4 h-\mu_p\kappa_0\partial_x^2\psi -\int_{q_x}e^{iq_xx}\frac{1}{4\tilde{\eta}q_x}\gamma_{act}q_x^2h
\label{Actpump1}
\end{equation}
\begin{equation}
\partial_t \psi= \tilde{v} \partial_x(\psi^2\partial_x h)+D\partial_x^2\psi-\kappa_0\partial_x^4 h-k_2\psi+k_1
\label{Actpump2}
\end{equation}
\label{Actpump}
\end{subequations}

\begin{figure}
 \begin{center}
  \subfigure{\includegraphics[height=2.0in]{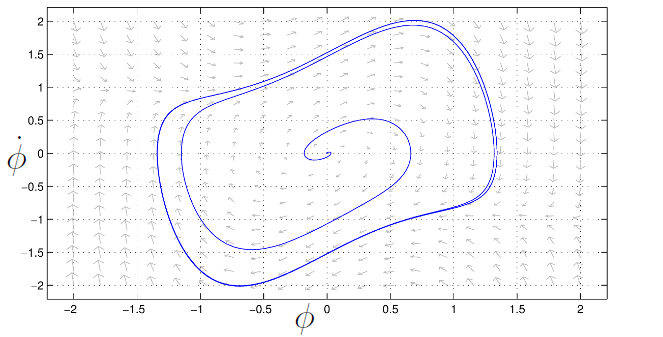}}
   \end{center}
\caption{A representative phase portrait showing a stable limit cycle. The
chosen parameter values are $s_1=-1$, $s_2=2.5$, $s_3=0.5$, and $s_4=1$. The
limit-cycle is formed through a supercritical Hopf bifurcation. \cite{pplane}}
\label{ph_port}
\end{figure}

The presence of the $\Sigma$ term will be argued for later in the paper. These equations with Neumann boundary conditions in an interval of length $\pi$, retaining only the smallest wavenumber ($q=1$), read

\begin{equation}
\frac{d}{dt} h_1= m_1 \psi_1-m_2 h_1
\label{h1t}
\end{equation}
\begin{equation}
\frac{d}{dt} \psi_1=-m_3(\psi_1\psi_1  h_1)- m_4\psi_1-m_5 h_1.
\label{p1t}
\end{equation}
Here, $m_1=\tilde{v}+\mu_p\kappa_0$, $m_2=\Sigma+\mu_p\kappa+\frac{\gamma_{act}}{4\eta}$; $m_3=\tilde{v}$; $m_4=D$; and $m_5=\kappa_0$.
These equations have eigenvalues
\begin{equation}
\frac{-(m_2+m_4)}{2}\pm \frac{[(m_2-m_4)^2-2m_1m_5]^{1/2}}{2}
\end{equation}
We can combine \eqref{h1t} and \eqref{p1t} into one second order differential equation, which upon defining $\phi=(m_5+m_2m_4)h$ and $\tau=\sqrt{(m_5+m_2m_4)}t$, yields
\begin{equation}
\ddot{\phi} +\phi+s_1\dot{\phi}+s_2 \phi^2\dot{\phi}+s_3 \phi
(\dot{\phi})^2 +s_4 \phi^3=0
\end{equation}
where an overdot denotes a derivative with respect to $\tau$. This is  a
generalised Van der Pol oscillator \cite{rand2} with a linear damping and cubic
nonlinearities, with new constants related to $m_i$ as follows:
\begin{equation}
s_1=\frac{m_2+m_4}{\sqrt{m_5+m_2m_4}}
\end{equation}
\begin{equation}
s_2=\frac{2m_3m_2}{m_1(m_5+m_2m_4)^{5/2}}
\end{equation}
\begin{equation}
s_3=\frac{m_3}{(m_5+m_2m_4)^2}
\end{equation}
\begin{equation}
s_4=\frac{m_3m_2^2}{m_1(m_5+m_2m_4)^3}
\end{equation}
We present a representative phase portrait in Fig. \ref{ph_port}for this equation which shows a limit cycle.

Note that this analysis is independent (except for the value of $m_2$) of
whether the membrane fluctuates in a fluid or with respect to a fixed background
medium. The main difference between the two cases is the absence, in the second
case, of a damping term linear in wavevector. In a mode-truncated description,
this detail does not change the conclusion about the presence of an
oscillatory state. 

\subsection{The $h-\psi$ excitatory-inhibitory coupling in an impermeable membrane}
The coupled eqs. (2) and (3) of the main text, in Monge gauge, imply that $\partial_t\psi$ will have a term proportional to $\nabla_{\perp}\cdot(\psi v_n\nabla_{\perp} h)$ due to kinematic reasons, where $v_n\equiv{ {\bf V}_m}\cdot {\bf N}$. In an impermeable membrane,  with no-slip in the tangential directions
\begin{equation}
{ {\bf V}_m}={\bf V}|_{{\bf r=R}}
\end{equation}
where ${\bf V}({\bf
r})=\int_{\bf r'} \bsf{H}({\bf r}-{\bf r'}) \cdot \nabla\cdot\bm{\sigma}({\bf r'})$, with $\bm{\sigma}$ being the hydrodynamic stress tensor. The lowest order polar active term in $\bm{\sigma}$ is $\zeta_p[\nabla {\bf P}+(\nabla {\bf P})^T]$. Integrating out ${\bf P}$ using eq. (5a) of the main text, we obtain 
\begin{equation}
v_n=-\int_{\bf
q}e^{i{\bf q}_{\perp}\cdot{\bf
x}}\frac{1}{4\tilde{\eta}q_{\perp}}\frac{\zeta_p\Lambda}{a_1}q_{\perp}^2\psi_{\bf q}
\end{equation}
from the polar stress. Thus, the kinematic term in the dynamical equation for $\psi$ is 
\begin{equation}
{ \frac{1}{4\tilde{\eta}q_{\perp}}}\frac{\zeta_p\Lambda}{a_1}\nabla_{\perp}\cdot(\psi\nabla_{\perp}h\int_{\bf q_{\perp}}e^{i{\bf q}_{\perp}\cdot{\bf x}} |q_{\perp}|\psi_{\bf q}).
\label{psieq1}
\end{equation}

Let us assume the system is in a noisy stationary state, either as a result of intability and chaos or because of the presence of thermal or chemical noise. We can then replace the pair of $\psi$ fields by their pair correlation in a kind of Hartree approximation. Thus, \eqref{psieq1} becomes
\begin{equation}
{ \frac{1}{4\tilde{\eta}q_{\perp}}}\frac{\zeta_p\Lambda}{a_1}\nabla_{\perp}\cdot(\nabla_{\perp}h\int_{\bf q_{\perp},{\bf k}}e^{i({\bf q}_{\perp}+{\bf k})\cdot{\bf x}}|q_{\perp}| <\psi_{\bf k}\psi_{\bf q}>)
\label{psiint}
\end{equation}
Defining $ <\psi_{\bf k}\psi_{\bf q}>=G_q\delta({\bf q}_{\perp}+{\bf k})$, we see that the integral in \eqref{psiint} reduces to $\int_{\bf q}|q_{\perp}|G_q$. Thus, as long as $G_q$ is not highly singular as $q_{\perp}\to 0$, we obtain waves with dispersion $\omega^2\sim q^3$ and speed $\zeta_p\Lambda/{(4\tilde{\eta}q_{\perp}a_1)}\sqrt{\int_{\bf q}|q_{\perp}|{ G_q}}$.

Note that the presence of the active polar stress is not a neccessary requirement for an active impermeable membrane to oscillate. If we consider an allowed free energy coupling $\psi\bsf{Q}_t\colon\bsf{K}$, where $\bsf{Q}_t={\bf e}^u{\bf e}^u\colon\bsf{Q}$ with ${\bf e}^u$ defined below eqn (1) of the main text, and integrate out $\bsf{Q}_t$ to obtain
\begin{equation}
\bsf{Q}_t\sim\int_u\psi\bsf{K},
\end{equation} we find 
\begin{equation}
v_n \sim-\int_{\bf
q}e^{i{\bf q}_{\perp}\cdot{\bf
x}}\frac{1}{4\tilde{\eta}q_{\perp}}q_{\perp}^2\psi_{\bf q}
\end{equation}
because of the active apolar stress. The argument presented in the previous paragraph, for spontaneously generated waves on the membrane, then follows through.

\subsection{Dispersion relations for coupled dynamics of ${\bf p}$ and $h$}
\begin{figure}
 \begin{center}
  \subfigure{\includegraphics[height=2.3in]{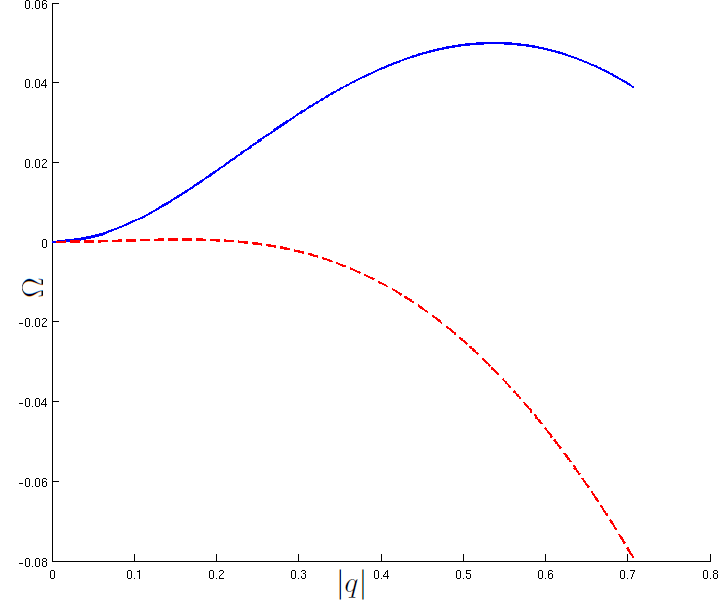}}
   \end{center}
\caption{Plot of the real part of the largest eigenvalue as a function of $|q|$ for $\alpha=-3.1$ (blue continuous line), and $\alpha=-2.1$ (red dashed line). All other parameter values are $1$}
\label{plt}
\end{figure}
Eqns.(11) in main text describe the coupled dynamics of membrane shape and concentration and polar orientation of horizontal actin filaments. Here we discuss the linear stability of this system. For $\tilde{A}>0$, i.e. when the orientation correlations of the polar filaments decay exponentially with distance, there are two modes with eigenvalues $\sim q^0$ and two with eigenvalues $\sim q^2$, unaffected, to leading order in wave-numbers, by the $\nabla_{\perp}\nabla_{\perp}^2 h$ coupling in ${\bf p}$. The absence of a linear coupling of $h$ to ${\bf p}$, at lower order in gradients, is ultimately a consequence of three-dimensional rotation invariance. Defining $\phi=\nabla_{\perp}\cdot{\bf p}$, the system of linearised equations in the isotropic phase is 
\begin{equation}
\partial_t\begin{pmatrix}h\\c\\ \phi\end{pmatrix}=\begin{pmatrix}-\Sigma q^2 & \tilde{v} &\zeta\Lambda_0/a_1 \\-\tilde{v}q^2 & -D_cq^2 & v_1\\0 & -\alpha q^2 & -\Gamma_p\tilde{A}\end{pmatrix}\begin{pmatrix}h\\c\\ \phi\end{pmatrix},
\end{equation}
where $\Lambda_0=\Lambda(c_0)$, $c_0$ being the mean concentration of the polar filaments. The eigenvalues, $\Omega$, of this set of equations are the solution of the equation
\begin{equation}
\Omega^3+m_1\Omega^2+m_2\Omega+m_3=0,
\end{equation}
\begin{equation}
\nonumber
m_1=(\Sigma+D_c)q^2+\Gamma_p\tilde{A}
\end{equation}
\begin{equation}
\nonumber
m_2=(\Sigma+D_c)\Gamma_p\tilde{A}q^2+\tilde{v}^2q^2+\alpha v_1q^2+D_c\Sigma q^4
\end{equation}
\begin{equation}
\nonumber
m_3=\tilde{v}^2\Gamma_p\tilde{A}q^2+D_c\Sigma\Gamma_p\tilde{A}q^4+\Sigma\alpha v_1q^4-\zeta\Lambda_0\tilde{v}\alpha/a_1q^4.
\end{equation}
with a general solution whose complicated form is not particularly enlightening. However, if other parameters are held fixed, we see that the eigenvalue with the maximum real part becomes positve as $\alpha v_1$ becomes large and negative (Fig. \ref{plt}). The corresponding eigenvector has a large projection on $h$, which increases with increasing value of $\tilde{v}$. The instability can be easily understood if the $\phi$ field is integrated out: it arises from the effective diffusivity in the $c$ equation becoming negative, and the resulting unbounded growth in local concentration promoting modulation of the membrane height.

We now provide the expressions of the eigenvalues of
the stability matrix. We will perturb about a state with a mean ${\bf p}_0=\sqrt{\frac{\tilde{A}}{\beta}}\hat{x}\equiv p_0\hat{x}$, where $\beta$ is the coefficient of the $P^4$ term in the free energy.
The transverse fluctuations of the order parameter, $p_y$, and the height field fluctuations are hydrodynamically slow in this phase. The general mode structure is quite complicated, with eigenvalues
\begin{widetext}
\begin{multline}
\Omega_{1,2}=\frac{1}{2}[-iv_p p_0 q_x-(\mu_p\Sigma+D_p)q^2-\kappa\mu_pq^4]\\\pm\frac{1}{2}[-b_4q_x^2+b_3q_x^2q_y^2+b_2q^2q_y^2+b_1q^4-b_6q^4q_y^2-b_5q^6+b_7q^8+i(b_8q_xq^2+b_9q_xq_y^2-b_{10}q_xq^4)]^{1/2}
\end{multline}
\end{widetext}
with the coefficients $b_i$ defined as
\begin{equation}
\nonumber
b_1=(D_p-\mu_p\Sigma)^2
\end{equation}
\begin{equation}
\nonumber
b_2=4\Gamma_p^{-1}\kappa_p\zeta\Lambda_0/a_1
\end{equation}
\begin{equation}
\nonumber
b_3=4\Gamma_p^{-1}\mu_p\kappa_t^2p_0^2
\end{equation}
\begin{equation}
\nonumber
b_4=v_p^2p_0^2
\end{equation}
\begin{equation}
\nonumber
b_5=2\kappa(D_p-\mu_p\Sigma)
\end{equation}
\begin{equation}
\nonumber
b_6=4\Gamma_p^{-1}\mu_p\kappa_p^2
\end{equation}
\begin{equation}
\nonumber
b_7=\kappa^2\mu_p^2
\end{equation}
\begin{equation}
\nonumber
b_8=2v_pp_0(D_p-\mu_p\Sigma)
\end{equation}
\begin{equation}
\nonumber
b_9=4\Gamma_p^{-1}\kappa_tp_0\zeta\Lambda_0/a_1
\end{equation}
\begin{equation}
\nonumber
b_{10}=2\kappa\mu_pv_pp_0
\end{equation}

\begin{figure}
 \begin{center}
  \subfigure{\includegraphics[height=1.8in]{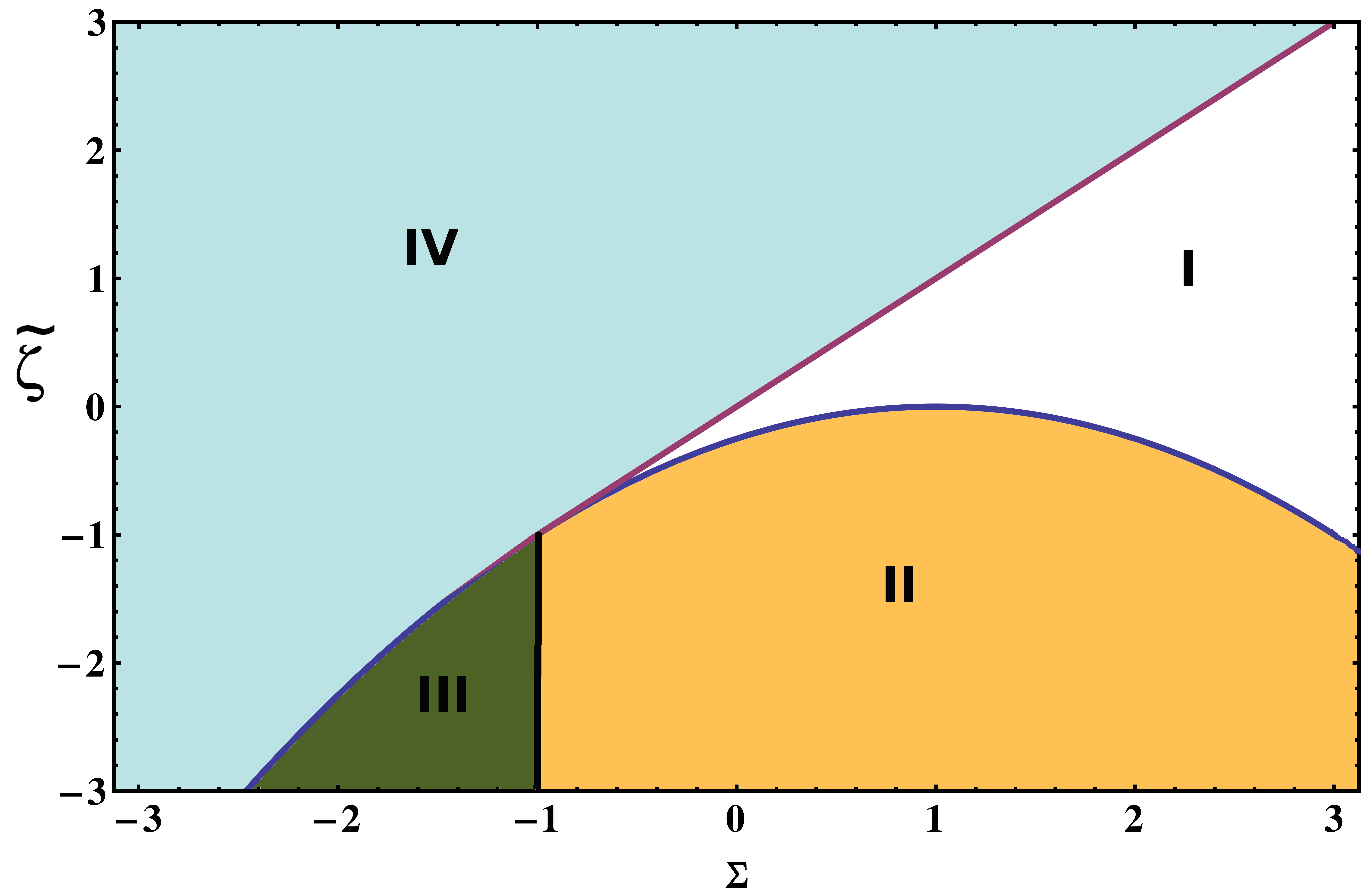}}
   \end{center}
\caption{Stability diagram for transverse fluctuations of a homogeneously polarized phase. {\bf I} : stable membrane, {\bf II} : decaying waves, {\bf III} : oscillatory instabilitiy and {\bf IV} : instability with $\sim q^2$ dispersion. }
\label{ph_dia}
\end{figure}

The mode structure for small wavevectors transverse to the ordering direction (i.e $q_x=0$, $q_y\to0$) is 
\begin{widetext}
\begin{equation}
\Omega_{1,2}=-\frac{1}{2}[(\mu_p\Sigma+D_p)\pm[(D_p-\mu_p\Sigma)^2+4\Gamma_p^{-1}\kappa_p\zeta\Lambda_0/a_1]^{1/2}]q_y^2-\frac{1}{2}\left[\kappa\mu_p\pm\frac{2\Gamma_p^{-1}\mu_p\kappa_p^2-\kappa(D_p-\mu_p\Sigma)}{[(D_p-\mu_p\Sigma)^2+4\Gamma_p^{-1}\kappa_p\zeta\Lambda_0/a_1]^{1/2}}\right]q_y^4
\end{equation}
\end{widetext}
We observe that these modes become propagative if $\zeta$ is large and negative. For $\zeta>0$ the ordered state becomes unstable even with a stabilising tension. We plot a stability diagram for this mode in Fig. \ref{ph_dia}

The mode structure for wavevectors along the ordering direction is
\begin{subequations}
\begin{equation}
\Omega_1=-iv_pp_0q_x-\frac{1}{2}(D_p+\mu_p\Sigma)q_x^2
\end{equation}
\begin{equation}
\Omega_2=-\frac{1}{2}(D_p+\mu_p\Sigma)q_x^2
\end{equation}
\end{subequations}

Now we examine the instability with the growth rate $q_x^{1/2}q_y$ as mentioned in the main text. This instability which arises from the combination of activity $\zeta$ and curvature coupling $\kappa_t$ [see equations (11) of main text], is best seen in the limit of immotile particles i.e. $v_p=0$. In this limit, the mode structure, to leading order in wavevectors, is
\begin{equation}
\Omega_{1,2}=\pm\frac{1+i\, sgn(\zeta\Lambda_0\kappa_tq_x)}{\sqrt{2}}\left(\frac{4\kappa_t\Gamma_p^{-1}p_0\zeta\Lambda_0}{a_1}\right)q_x^{1/2}q_y
\end{equation}
Since $\Omega_{1,2}$ have real and imaginary parts, the mode displaying the instability travels in a direction governed by sgn$(\zeta\Lambda_0\kappa_t)$.

If the particles are slightly motile, i.e. with $v_p$ which is smaller than the other velocity scales in the problem, the mode structure, to leading order in gradients and $v_p$ is
\begin{subequations}
\begin{equation}
\Omega_1=2\frac{\kappa_t\zeta\Lambda_0}{\Gamma_pv_pa_1}q_y^2
\end{equation}
\begin{equation}
\Omega_2=-i v_pp_0 q_x-2\frac{\kappa_t\zeta\Lambda_0}{\Gamma_pv_pa_1}q_y^2
\end{equation}
\end{subequations}
We see that the instability is weakened with increasing motility, another feature in common with \cite{Sumithra1}. 






\begin{thebibliography}{}
\bibitem{Sumithra} No substrate is present to provide a fixed reference
horizontal, unlike in \cite{Sumithra2}.
\bibitem{conc} We ignore the fluctuations of the concentration of the filaments
in the bulk fluid.
\bibitem{Nir}  There can be no monopole term in the force density because of
momentum conservation.  Such a term is introduced in \cite{gov}, but its
presence is thereafter rendered innocuous once the authors replace hydrodynamic
damping by local friction.
\bibitem{Propulsion} We ignore other possible terms at this order in gradients
and fields \cite{Toner} which do not play a significant role
here.
\end{thebibliography}
\end{document}